\documentclass[11pt, a4paper, logo, copyright, nonumbering]{Bohr_Sci}

\usepackage{graphicx}%
\usepackage{multirow}%
\usepackage{amsmath,amssymb,amsfonts}%
\usepackage{amsthm}%
\usepackage{mathrsfs}%
\usepackage[title]{appendix}%
\usepackage{xcolor}%
\usepackage{textcomp}%
\usepackage{manyfoot}%
\usepackage{booktabs}%
\usepackage{algorithm}%
\usepackage{algorithmicx}%
\usepackage{algpseudocode}%
\usepackage{listings}%
\usepackage{mathtools}
\usepackage{tabularx}
\usepackage[most]{tcolorbox}
\usepackage{letltxmacro}
\usepackage{xstring}
\usepackage{environ}
\usepackage{array}

\usepackage[most]{tcolorbox}
\usepackage{enumitem}

\definecolor{agentbg}{RGB}{245,248,252}      
\definecolor{agentborder}{RGB}{210,220,235}  
\definecolor{agenttitle}{RGB}{40,60,110}     

\newtcolorbox{agentcard}[1][]{
  enhanced,
  breakable,
  colback=agentbg,
  colframe=agentborder,
  coltitle=agenttitle,
  boxrule=0.6pt,
  arc=3pt,
  left=6pt,right=6pt,top=6pt,bottom=6pt,
  title={#1},
  fonttitle=\bfseries\large,
}

\usepackage{tikz}
\usetikzlibrary{arrows.meta, positioning, fit, shapes.geometric, shapes.misc}

\theoremstyle{thmstyleone}%
\theoremstyle{thmstyletwo}%

\theoremstyle{thmstylethree}%

\usepackage[capitalize,noabbrev]{cleveref}

\definecolor{wikiblue}{HTML}{1195DB}
\definecolor{Grokipediaorange}{HTML}{F4A324} 
\definecolor{sciencepediagreen}{HTML}{27B57A}
\definecolor{baselinecolor}{HTML}{DC5151}

\newtcolorbox{wikipediaarticle}[1][Wikipedia Article]{
    colback=wikiblue!10!white,
    colframe=wikiblue,
    fonttitle=\bfseries,
    breakable,
    title={#1}
}

\newtcolorbox{sciencepediaarticle}[1][Sciencepedia Article]{
    colback=sciencepediagreen!10!white,
    colframe=sciencepediagreen,
    fonttitle=\bfseries,
    breakable,
    title={#1}
}

\newtcolorbox{llmbaseline}[1][Baseline LLM Article]{
    colback=baselinecolor!10!white,
    colframe=baselinecolor,
    fonttitle=\bfseries,
    breakable,
    title={#1}
}

\newtcolorbox{grokipediaarticle}[1][Grokipedia Article]{
    colback=Grokipediaorange!10!white,
    colframe=Grokipediaorange,
    fonttitle=\bfseries,
    breakable,
    title={#1}
}

\title{\textsc{Deploy-Master}: Automating the Deployment of 50,000+ Agent-Ready Scientific Tools in One Day}

\author{
Yi Wang\textsuperscript{1}
Zhenting Huang\textsuperscript{1}
Zhaohan Ding\textsuperscript{1,$\dagger$}
Ruoxue Liao\textsuperscript{1}
Yuan Huang\textsuperscript{1}
Xinzijian Liu\textsuperscript{1}
Jiajun Xie\textsuperscript{1}
Siheng Chen\textsuperscript{3}
Linfeng Zhang\textsuperscript{1,2,$\dagger$}
\\
\textsuperscript{1} DP Technology, Beijing, China \\
\textsuperscript{2} AI for Science Institute, Beijing, China \\
\textsuperscript{3} Shanghai Jiao Tong University, Shanghai, China \\
\normalfont{$\dagger$ Corresponding authors: dingzh@dp.tech; zhanglf@dp.tech}
}

\begin{document}

\maketitle
\thispagestyle{firststyle}

\definecolor{Scimaster_Blue}{HTML}{004098}
\centerline{\color{Scimaster_Blue}\fontsize{15pt}{14pt}\selectfont\textbf{Abstract}}
\vspace{2ex}

\absfont
\noindent{Open-source scientific software is abundant, yet most tools remain difficult to compile, configure, and reuse, sustaining a small-workshop mode of scientific computing. This deployment bottleneck limits reproducibility, large-scale evaluation, and the practical integration of scientific tools into modern AI-for-Science (AI4S) and agentic workflows.}

\noindent{We present \textsc{Deploy-Master}, a one-stop agentic workflow for large-scale tool discovery, build specification inference, execution-based validation, and publication. Guided by a taxonomy spanning 90+ scientific and engineering domains, our discovery stage starts from a recall-oriented pool of over 500,000 public repositories and progressively filters it to 52,550 executable tool candidates under license- and quality-aware criteria. \textsc{Deploy-Master} transforms heterogeneous open-source repositories into runnable, containerized capabilities grounded in execution rather than documentation claims. In a single day, we performed 52,550 build attempts and constructed reproducible runtime environments for 50,112 scientific tools. Each successful tool is validated by a minimal executable command and registered in \textsc{SciencePedia} for search and reuse, enabling direct human use and optional agent-based invocation.}

\noindent{Beyond delivering runnable tools, we report a deployment trace at the scale of 50,000 tools, characterizing throughput, cost profiles, failure surfaces, and specification uncertainty that become visible only at scale. These results explain why scientific software remains difficult to operationalize and motivate shared, observable execution substrates as a foundation for scalable AI4S and agentic science.}


\section{Introduction}
\label{sec:introduction}
Over the past decades, scientific computing has produced an unprecedented number of open-source software tools across nearly all disciplines, ranging from bioinformatics and chemistry to materials science, physics, and engineering. Despite this abundance, the \emph{usability} of scientific software has not scaled accordingly. Most tools remain difficult to compile, configure, and reproduce, relying heavily on manual intervention, undocumented assumptions, and environment-specific expertise \cite{wilson2014best,sandve2013ten}. As a result, a large fraction of scientific software exists in repositories and publications, but not in a form that can be readily executed, shared, or reused.

This long-standing gap can be characterized as a \emph{small-workshop mode} of scientific software deployment: individual researchers or groups hand-craft their own environments, often spending days or weeks resolving dependencies and compilation issues \cite{wilson2014best}. These environments are typically non-portable, non-reproducible, and difficult to share even within the same institution, let alone across computing platforms. While containerization, cloud computing, and HPC infrastructures have significantly improved resource accessibility, they have not fundamentally eliminated the deployment bottleneck \cite{kurtzer2017singularity}. In practice, the effort required to make a scientific tool runnable remains a major barrier to reuse, large-scale evaluation, and systematic integration into modern scientific workflows. Moreover, as tool diversity and volume continue to grow, the primary friction is often not that tools are insufficiently accurate or efficient, but that the \emph{human migration cost}---installation, environment debugging, interface conventions, and workflow rewiring---is too high for widespread adoption. Software that is even 2--3$\times$ better may still fail to diffuse if switching remains expensive.

This bottleneck has become increasingly problematic with the rise of AI for Science (AI4S). Contemporary AI4S systems are no longer centered solely on models or algorithms; they increasingly depend on the tight integration of models, scientific tools, workflows, and execution environments. Models are expected not only to make predictions, but also to invoke external solvers, simulators, data processors, and analysis pipelines \cite{yao2022react}. In this context, whether a tool is \emph{actually runnable} becomes a first-order concern: a tool that cannot be executed reliably cannot be meaningfully incorporated into AI-assisted workflows, benchmarked at scale, or compared across tasks and domains. More broadly, when the available tool set grows from hundreds to tens of thousands, the interaction model shifts: tools are no longer primarily integrated ``one by one'' by human experts, but must become callable capabilities that can be composed, tested, and selected under execution-grounded criteria.

The same issue appears even more sharply in the emerging paradigm of \emph{agentic science at scale}. Scientific agents are expected to plan, select, and execute tools autonomously, iterating over hypotheses and experiments. However, tool invocation in such systems is only as reliable as the underlying execution substrate. When tools are fragile, undocumented, or environment-dependent, planning cannot be grounded in execution, failures cannot be systematically analyzed, and execution traces cannot be reused as learning signals \cite{zhang2025bohriumscimaster}. In this sense, tools that are not deployment-ready form a structural bottleneck for scaling agentic scientific systems. Importantly, this limitation affects not only autonomous agents, but also human researchers who wish to reliably run, reproduce, and compare scientific software at scale. Conversely, for agents that can amortize switching costs through repeated trials and automated invocation, making tools reliably runnable can unlock a much larger design space of ``better-but-harder-to-adopt'' methods.

In this work, we address this problem from an infrastructure perspective. We present \textsc{Deploy-Master}, an \emph{agentic workflow} for large-scale discovery, compilation, deployment, and publication of open-source scientific tools. Concretely, a user provides a repository URL, and Deploy-Master automatically infers build specifications, constructs a reproducible runtime image, validates it via a minimal executable command, and publishes the resulting runnable capability.\footnote{\url{https://www.bohrium.com/apps/deploy-master} \ \textit{(Deploy-Master web entry)}} Using Deploy-Master, we performed 52,550 build attempts across open-source repositories (primarily from GitHub) and successfully constructed runnable environments for 50,112 scientific tools within a single day. From a user perspective, Deploy-Master provides a direct path from ``a repository that exists'' to ``a tool that runs'': the output can be executed interactively (e.g., notebooks) or at scale (e.g., batch jobs) without manual environment reconstruction.

The resulting tools are published through a unified interface as part of \emph{SciencePedia}, a cross-disciplinary hub that indexes execution-validated scientific tools with structured metadata (e.g., descriptions, tags, and entrypoints) for search and reuse.\footnote{\url{https://sciencepedia.bohrium.com} \ \textit{(SciencePedia tool hub)}} These artifacts support direct human use---including batch execution, interactive notebooks, and scalable cloud/HPC workloads---without requiring users to manually reconstruct environments. Optionally, Deploy-Master can expose tools through standardized agent-facing interfaces (e.g., MCP endpoints), enabling agents to invoke tools as registered and traceable actions within larger workflows \cite{mcp2025spec}. Rather than treating agent-readiness as a prerequisite, Deploy-Master treats it as a natural extension of making tools reliably runnable.

Beyond delivering a large corpus of usable scientific tools, this work provides an empirical lens on deployability at scale. By analyzing throughput, build-time cost profiles, failure surfaces, and specification uncertainty across tens of thousands of real repositories, we surface systematic signals that are difficult to observe at smaller scales. These results offer concrete insights into why scientific software remains hard to operationalize and what infrastructure-level interventions are required to support scalable AI4S and agentic science. Finally, Deploy-Master is intended to serve as a capability-conversion layer within a broader ecosystem stack: execution-validated tools can be composed by \emph{SciMaster}-style orchestrators and hierarchies of master agents and community agents, while large-scale executions generate observable traces that feed back into build specification refinement, tool ranking, and workflow planning \cite{zhang2025bohriumscimaster}.

In summary, this paper makes three contributions. First, we demonstrate that tens of thousands of heterogeneous scientific tools can be automatically transformed into runnable, reusable capabilities within practical time constraints. Second, we present Deploy-Master, an agentic workflow that integrates tool discovery, automated build specification inference, execution validation, and publication into a single scalable pipeline. Third, we provide large-scale empirical evidence characterizing the deployment landscape of scientific software, highlighting key challenges and opportunities for building shared, executable substrates for AI4S and agentic science at scale.

\section{Deploy-Master System Overview}
\label{sec:system}
Deploy-Master is designed as a one-stop agentic workflow that transforms open-source scientific software repositories into runnable and reusable capabilities at scale. Rather than focusing on individual tools or handcrafted environments, the system integrates large-scale tool discovery, automated build specification inference, execution validation, and publication into a unified pipeline. This section provides a system-level overview of Deploy-Master, emphasizing what the system does and what artifacts it produces, rather than low-level implementation details.

\subsection{Tool Discovery at Scale (Search Agent)}
\label{subsec:search-agent}

A central challenge in large-scale scientific tool discovery is balancing coverage and precision. Queries that are too broad retrieve large volumes of irrelevant repositories, while overly narrow queries miss substantial portions of the scientific tool ecosystem. This trade-off makes naive keyword-based search insufficient for systematic discovery at scale.

Deploy-Master addresses this challenge by grounding discovery in a structured organization of the AI4S landscape that reflects how scientific software is used in practice. Concretely, we start from a taxonomy of 91 scientific and engineering domains and use a language model to expand each domain into a small set of core search keywords. These keywords drive large-scale retrieval over GitHub and the public web, producing an initial, recall-oriented candidate pool of more than 500{,}000 repositories.

To improve recall while maintaining relevance, repositories identified in the initial search are used as anchors for iterative expansion. This expansion leverages signals such as dependency relationships, referenced tools, shared contributors, and linked documentation to identify related repositories. Repository metadata are further enriched to extract programming languages, documentation availability, and basic structural properties relevant to executability. Lightweight heuristic filters are then applied to remove repositories that are unlikely to represent usable scientific tools, such as documentation-only repositories, curated lists, tutorials, and projects lacking executable content.

\paragraph{Multi-Stage Candidate Funnel.}
At the scale required for this work, tool discovery is best treated as a progressive filtering process that trades recall for precision across multiple stages. Starting from taxonomy-guided keyword search over GitHub and the public web, Deploy-Master initially retrieved more than 500{,}000 candidate repositories. After lightweight heuristic filtering and de-duplication, the candidate set was reduced to 240{,}645 tool-like repositories. In the final stage, an agent-based semantic filtering process identified repositories that qualify as executable scientific tools; after this stage, 52{,}550 repositories were retained as final tool candidates and passed to the automated build and validation pipeline.

\begin{figure}[!t]
    \centering
    \includegraphics[width=\linewidth]{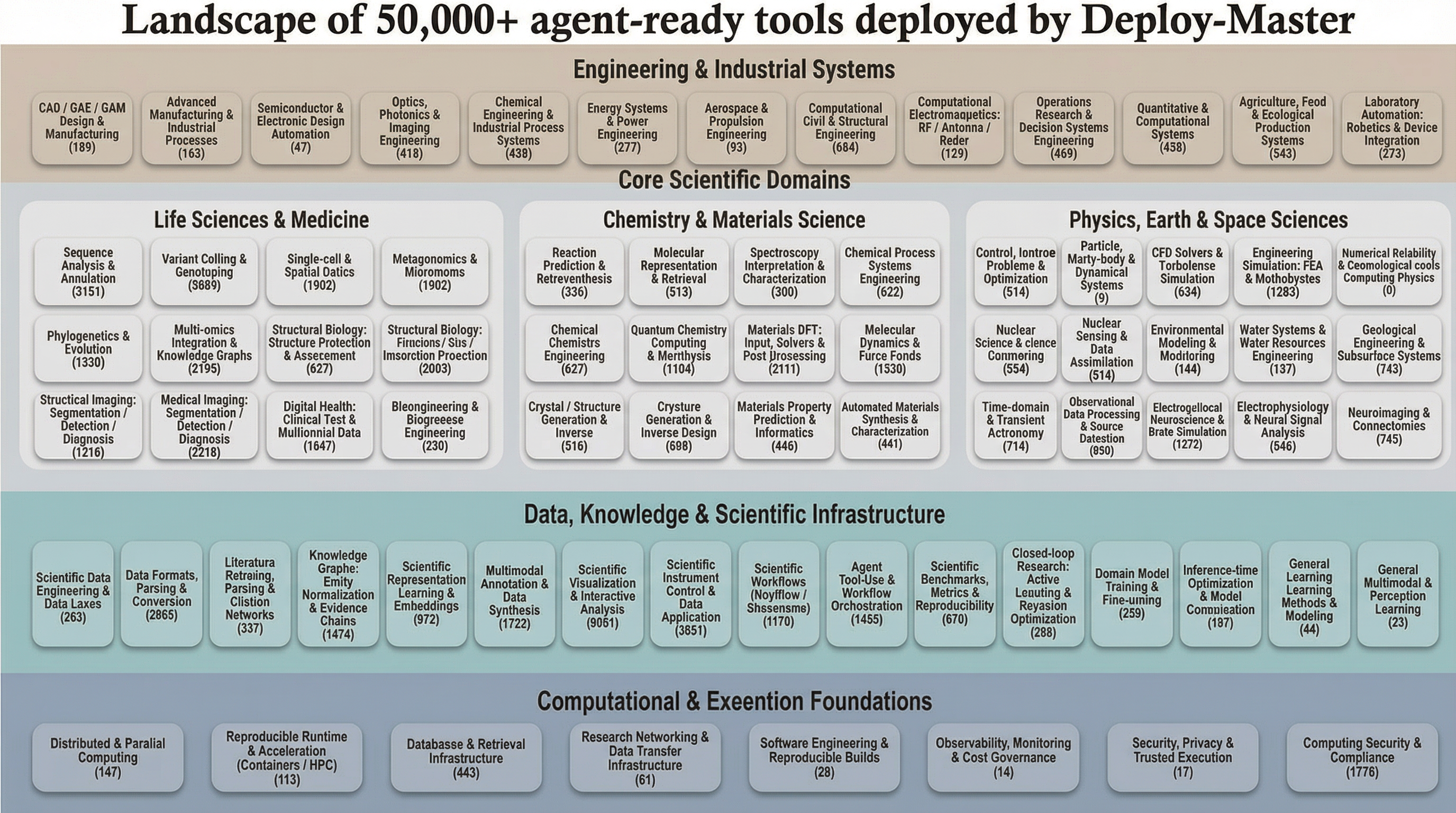}
    \caption{Landscape of scientific tools constructed in this work, organized by scientific and engineering application domains. Domain assignment is performed by embedding-based similarity matching between tool descriptions and domain definitions; the visualization uses a similarity threshold of 0.5. Tools may belong to multiple categories due to overlapping functionalities across tasks and disciplines; therefore, the summed counts across categories exceed the total number of unique tools.}
    \label{fig:tool-landscape}
\end{figure}

After successful builds are produced, we associate tools back to the 91-domain taxonomy via embedding-based similarity matching between tool descriptions and domain definitions. Figure~\ref{fig:tool-landscape} visualizes the resulting tool landscape using a similarity threshold of 0.5. Because many tools naturally support multiple tasks or domains, categories are not mutually exclusive, and aggregated counts across categories exceed the total number of unique tools.

\subsection{Automated Build and Validation Pipeline (Build Agent)}
\label{subsec:build-agent}

Given a candidate repository, Deploy-Master employs an automated build and validation pipeline to infer how the tool can be executed and to construct a reproducible runtime environment. The pipeline is fully automated and explicitly designed to operate under incomplete, inconsistent, or outdated documentation, which is common across real-world scientific repositories.

At a high level, the pipeline consists of the following stages:
(1) repository cloning and structural analysis;
(2) systematic traversal and parsing of build- and run-related artifacts within the repository, including README files, INSTALL guides, auxiliary documentation, existing Dockerfiles, and workflow specifications such as GitHub Actions;
(3) supplemental web search to retrieve installation instructions and dependency information when repository-provided materials are insufficient;
(4) automatic generation of a candidate build specification (e.g., a Dockerfile);
(5) dual-model evaluation and debate-based refinement of the build specification;
(6) container image construction;
(7) execution validation via a minimal runnable command; and
(8) publication and registration of successful builds.

This design is motivated by a practical observation shared with prior reproducible-environment systems: repositories often contain partial signals (e.g., dependency files or build scripts), but these signals are rarely sufficient as a complete executable specification at scale \cite{forde2018repo2docker，bussonnier2018binder2}. Deploy-Master therefore treats build specifications as hypotheses assembled from heterogeneous evidence, then verified through execution.

A key design feature of Deploy-Master is the use of a dual-model evaluation loop during build specification generation. Initial specifications produced by a single model frequently fail due to mismatches between documentation and repository state, missing dependencies, or implicit environment assumptions. To mitigate this, a second model independently reviews the generated specification against repository artifacts and supplemental information, identifies inconsistencies, and proposes corrections. The two models iteratively refine the specification through mutual evaluation until a stable configuration is obtained.

After image construction, each tool is validated using a minimal executable command that exercises its primary entry point. This validation step establishes an execution-based criterion for usability: a tool is considered successfully built if it can be executed in the constructed environment without manual intervention. Successful builds are then registered, annotated with structured metadata (e.g., descriptions, tags, and entrypoints), and published through a unified interface, making them available for direct human use or downstream integration. At a high level, this publication step aligns with the broader goal of ecosystem-level tool dissemination pursued by community distribution efforts (e.g., packaging and container registries), while focusing on execution-validated capabilities and cross-domain coverage \cite{gruning2018bioconda,leprevost2017biocontainers}.

\begin{figure}[!t]
    \centering
    \includegraphics[width=\linewidth]{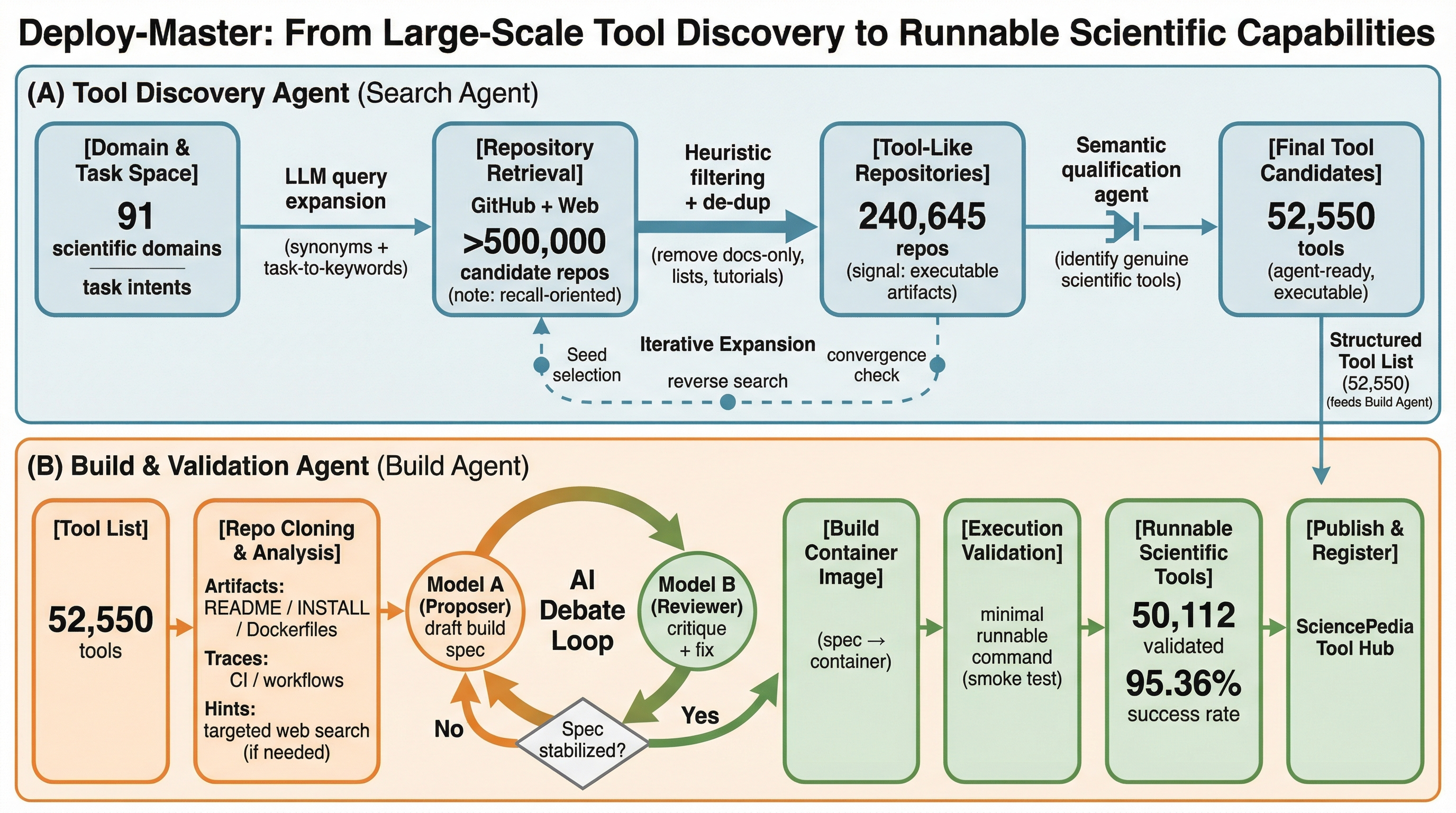}
    \caption{Deploy-Master workflow from large-scale discovery to runnable scientific capabilities. \textbf{(A) Search Agent:} taxonomy-guided retrieval (91 domains) and iterative expansion produce a recall-oriented pool ($>$500{,}000), which is progressively filtered to 240{,}645 tool-like repositories and then to 52{,}550 executable tool candidates. \textbf{(B) Build Agent:} each candidate is processed via repository analysis, build-specification proposal, dual-model evaluation and refinement, container construction, and execution validation, yielding 50{,}112 execution-validated tools (95.36\% success rate) that are published and registered for search and reuse.}
    \label{fig:deploy-master-pipeline}
\end{figure}

The overall discovery-to-publication workflow is summarized in Figure~\ref{fig:deploy-master-pipeline}.

\subsection{Execution-Centered Deployment Under Imperfect Information}
\label{subsec:not-better-dockerfiles}

A central challenge in large-scale scientific software deployment is not containerization itself, but the gap between incomplete specifications and executable reality. Repository documents and build recipes are often partial, outdated, or inconsistent with the current codebase, while critical assumptions about operating systems, compilers, and system libraries remain implicit. At the scale of tens of thousands of tools, such mismatches dominate behavior and make purely documentation-driven automation brittle.

In early experiments, generating Dockerfiles with a single model based primarily on repository documentation led to build success rates of only 50--60\%. The dominant source of failure was not container technology, but pervasive mismatch between documentation, repository structure, and actual build requirements. README files are frequently stale, existing Dockerfiles may no longer reflect current dependencies, and base image choices can silently invalidate installation steps (e.g., due to end-of-life distributions or broken package repositories).

Deploy-Master addresses this challenge by treating build specifications as hypotheses that must be evaluated and refined. The dual-model evaluation and debate mechanism reduces fragile assumptions by independently reviewing, critiquing, and revising the build specification against repository artifacts and supplemental information. This iterative process substantially improves robustness, enabling build success rates above 95\% across tens of thousands of tools.

More broadly, Deploy-Master reframes deployment as the production of \emph{runnable facts} under imperfect information. By grounding usability in execution rather than documentation claims, the system provides a more reliable execution substrate for large-scale AI4S workflows and agentic scientific systems, where stable tool invocation is a prerequisite for planning, execution, and learning. When exposed via standardized agent-facing interfaces (e.g., MCP endpoints), each tool becomes a registered, traceable action that can be invoked by agents as part of larger workflows \cite{mcp2025spec}.

\section{Results and Discussion: A Large-Scale Deployment Trace of 50,000+ Scientific Tools}
\label{sec:results}
This section reports the empirical outcomes of deploying scientific software at scale using Deploy-Master. Rather than seeking to isolate causal factors or derive prescriptive rules, we present a deployment trace that captures what occurs when tens of thousands of heterogeneous repositories are converted into execution-validated tools under shared infrastructure constraints. Our goal is to surface quantitative facts and engineering signals that are difficult to observe at smaller scales, and to discuss what they imply for building reliable execution substrates that can support AI4S pipelines and agentic scientific systems.

\subsection{One-Stop Deployment Outcomes and Cost Profile}
\label{subsec:overall-results}

We first examine the aggregate outcomes of one-stop deployment, from repository intake to execution-validated publication. Using Deploy-Master, we conducted 52,550 build attempts across open-source scientific repositories. Among these attempts, 50,112 tools were successfully built and validated, while 2,438 failed, yielding an overall success rate of 95.36\%. This outcome reflects a stabilized deployment regime after iterative improvements to scheduling, resource management, and specification refinement.

\begin{figure*}[t]
    \centering
    \includegraphics[width=\linewidth]{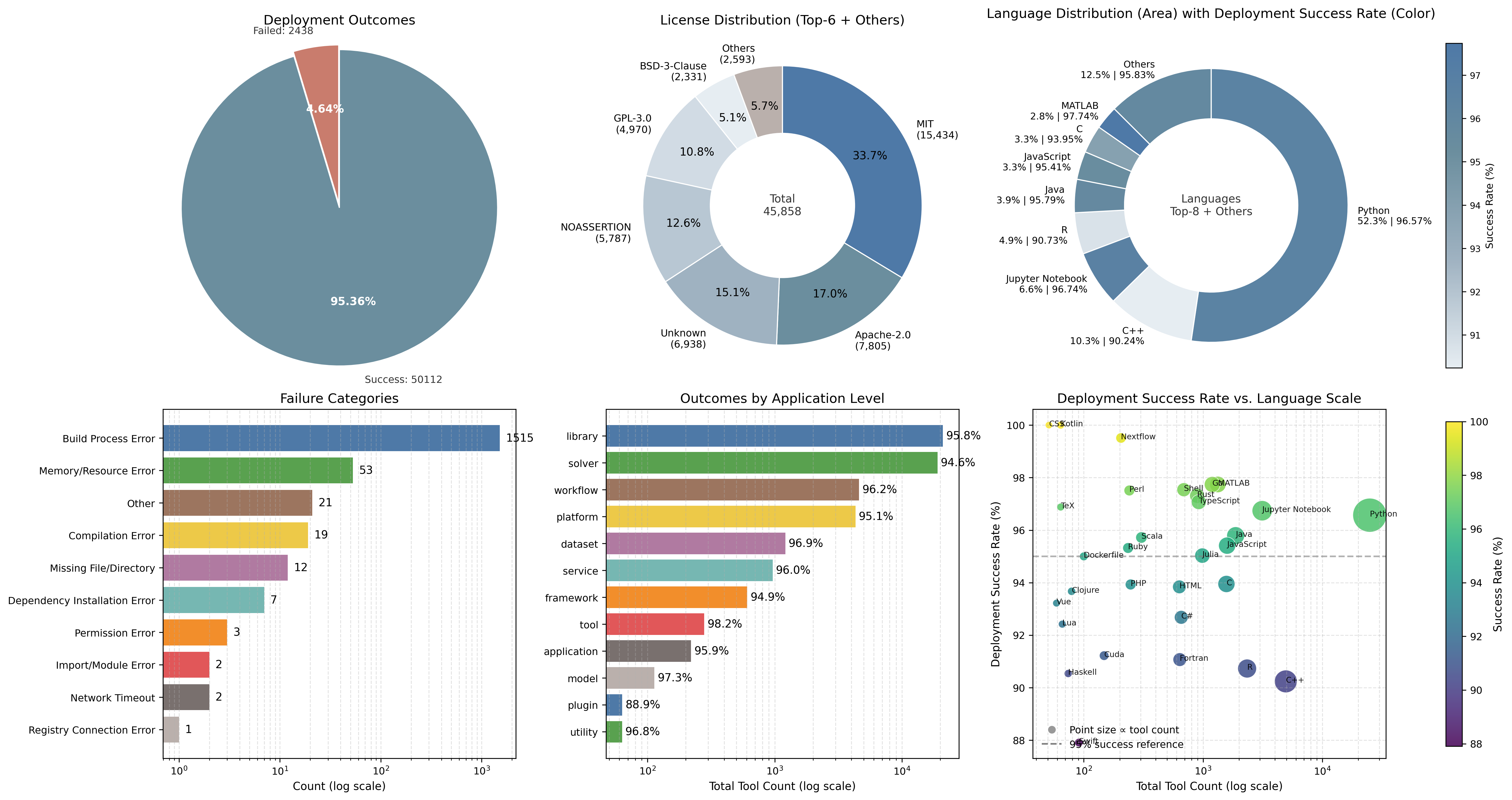}
    \caption{Summary of large-scale deployment outcomes and corpus characteristics. The figure aggregates (top left) overall deployment success and failure rates, (top middle) license distribution, (top right) programming language distribution with deployment success rates, (bottom left) failure categories, (bottom middle) outcomes by application level, and (bottom right) deployment success rate versus language scale.}
    \label{fig:deployment-summary}
\end{figure*}

Figure~\ref{fig:deployment-summary} provides a consolidated overview of deployment outcomes, corpus composition, and observed failure surfaces. The aggregate success rate indicates that automated deployment over a highly heterogeneous tool corpus can be sustained at the scale of tens of thousands under a unified, execution-centered workflow, while the remaining failures expose the dominant uncertainty and cost drivers that the infrastructure must manage.

Build time provides a complementary view of scalability. As shown in earlier analyses, while the median build time remains under ten minutes, the distribution exhibits a pronounced long tail. This long-tail behavior reflects substantial variation in build complexity, ranging from lightweight scripts to larger systems with extensive compilation steps and transitive dependencies. Although such variability does not prevent sustained throughput at scale, it shapes the cost profile that deployment infrastructure must absorb, reframing large-scale deployment as an execution-and-resource management problem in addition to a specification inference problem. In particular, long-tail builds create backpressure under high concurrency and motivate scheduling and isolation policies that are sensitive to both expected latency and resource footprint.

Finally, the deployed corpus is intrinsically diverse. The successfully built tools span more than 170 programming languages, with a small number of dominant languages accounting for a large fraction of repositories and a long tail of languages appearing infrequently. This diversity provides important context: Deploy-Master operates over a heterogeneous workload in which build conventions, runtime assumptions, and levels of specification explicitness vary widely across ecosystems.

\subsection{Corpus Composition, Language Scale, and Deployment Robustness}
\label{subsec:corpus-composition}

To further contextualize deployment behavior, we examine corpus composition through the lens of programming languages and tool scale. Programming language usage is highly skewed: Python alone accounts for over half of all successfully deployed tools, followed by C/C++, notebook-based tools, R, and Java, with dozens of additional languages appearing at much smaller scales.

Despite this skew, deployment success rates remain high across most languages. As illustrated in Figure~\ref{fig:deployment-summary}, the majority of languages cluster around success rates above 95\%, even as tool counts span several orders of magnitude. A small subset of languages exhibits comparatively lower success rates, including C/C++, Fortran, and R, as well as several low-frequency, system-oriented languages.

Importantly, these differences should not be interpreted as intrinsic properties of the languages themselves. Rather, lower success rates correlate with build characteristics that amplify specification uncertainty: compiled toolchains, reliance on system-level libraries (e.g., BLAS, MPI, CUDA), and implicit assumptions about operating systems and compiler environments. In this sense, observed variation in success rate serves as a proxy for environmental coupling and build-time uncertainty, rather than language-level ``deployability.''

Even for these more challenging cases, success rates remain close to or above 90\%, indicating that execution-centered validation combined with iterative specification refinement can substantially compress the deployability gap between lightweight, interpreted tools and heavily coupled, compiled systems. At scale, language heterogeneity therefore defines the workload distribution faced by the deployment system, rather than imposing a hard boundary on what can be made runnable.

\subsection{Failure Surface and Debugging Signals at Scale}
\label{subsec:failure-surface}

Failures at scale reveal where automated deployment encounters uncertainty and friction. Across unsuccessful build attempts, failure modes are highly non-uniform. As shown in Figure~\ref{fig:deployment-summary}, build process errors dominate the failure surface by a wide margin, far exceeding resource-related errors, dependency installation failures, permission issues, or network-related problems.

Build process errors include cases where build instructions are incomplete, inconsistent with the repository state, or incompatible with the execution environment. These failures are defined strictly at the level of observable execution outcomes, such as compiler errors or non-zero exit states, rather than inferred root causes. Resource-related errors, by contrast, primarily reflect contention effects under high concurrency, including memory exhaustion and disk pressure.

Rather than treating failures as static properties of tools, we interpret them as structured debugging signals for the deployment system itself. For example, the prevalence of build process errors highlights the central role of specification uncertainty in scientific software, while early observations of resource-related failures motivated subsequent refinements in scheduling and isolation strategies. Similarly, differences in failure profiles across language ecosystems reflect variation in how much information must be inferred during build specification generation, rather than categorical incompatibilities.

By treating failures as observable signals rather than terminal outcomes, large-scale deployment shifts from a binary notion of success versus failure toward an iterative engineering process. This perspective is essential when operating under fixed resource budgets and high concurrency, and it also enables downstream agentic systems to reuse execution traces as learning signals---e.g., to avoid brittle tools, to prioritize robust capabilities, or to trigger targeted remediation for high-value failures.

\subsection{Build-System Artifacts as a Scalability Proxy}
\label{subsec:build-artifacts}

A practical challenge in automated deployment is estimating deployability using signals that are both machine-readable and available at scale. As a proxy for specification availability, we analyze the presence of explicit build-system artifacts within repositories, such as \texttt{setup.py}, \texttt{pyproject.toml}, \texttt{requirements.txt}, \texttt{Makefile}, and \texttt{CMakeLists.txt} \cite{forde2018repo2docker}.

Across the corpus, a substantial fraction of repositories lack any explicit build-system artifacts of this form. Others provide only a single configuration file, while a smaller subset includes multiple build-related specifications. This distribution highlights a key source of deployment uncertainty: in the absence of explicit, machine-readable instructions, the system must infer execution requirements from incomplete or implicit signals.

Rather than asserting that missing build artifacts directly cause failure, we interpret this pattern as a measure of specification uncertainty. Repositories with limited or absent build metadata expand the search space for viable execution environments, motivating the design choices described in Section~\ref{sec:system}, including comprehensive repository traversal, supplemental information retrieval, and iterative refinement through model-based evaluation. At scale, such proxies offer actionable guidance for system design by indicating where robustness mechanisms are most needed to reduce uncertainty at the tool-execution layer and to improve the reliability of the resulting capability space.

\section{Outlook: Toward a Shared Execution Substrate for Agentic Science at Scale}
\label{sec:outlook}
The results in Section~\ref{sec:results} show that large-scale deployment of scientific software is not only feasible, but also diagnostic: once deployment is executed at the scale of tens of thousands of tools, the dominant frictions become measurable and actionable. Beyond the immediate empirical findings, this raises a broader question: if deployability is a structural bottleneck for AI4S and agentic science, what infrastructure is required beyond making individual tools runnable? In this section, we outline how Deploy-Master fits into a larger trajectory toward shared, reliable execution substrates for scientific intelligence, while remaining explicit about current limitations \cite{zhang2025bohriumscimaster}.

\subsection{From Runnable Tools to Shared Scientific Environments}
\label{subsec:shared-environments}

Deploy-Master addresses a foundational but deliberately scoped problem: transforming heterogeneous scientific software repositories into runnable, reproducible capabilities grounded in execution rather than documentation. This converts ``software that exists'' into ``tools that run,'' providing a concrete substrate that both human researchers and automated systems can invoke reliably.

However, runnable tools alone do not constitute a complete scientific environment. As tool availability grows into the tens of thousands, isolated container images become insufficient for sustained workflows. At this scale, bottlenecks shift toward shared runtime management, resource isolation, execution observability, and governance. In other words, once tool execution becomes possible, the limiting factor increasingly becomes how execution is orchestrated, monitored, and controlled across many tools, users, and workloads.

From this perspective, Deploy-Master is a building block toward shared scientific environments rather than an end solution. By standardizing build, validation, and registration, it enables higher-level systems to reason about execution in a uniform way and to treat tools as managed capabilities rather than ad hoc codebases. Importantly, large-scale executions also produce structured traces (success/failure states, resource footprints, and environment deltas) that can be aggregated into operational signals for improving scheduling policies, build specification refinement, and tool metadata quality. Integrating such capabilities into shared execution platforms can therefore support reproducible experiments, controlled resource usage, and systematic analysis of execution outcomes across tools and domains.

\subsection{Implications for Agentic Science and AI4S Evaluation}
\label{subsec:agentic-implications}

These results are particularly salient for agentic science and the evaluation of AI4S systems. Autonomous scientific agents are often discussed in terms of planning and reasoning, but their effectiveness ultimately depends on reliable interaction with external tools. In tool-rich but execution-poor settings---where many tools exist but invocation is fragile, undocumented, or environment-dependent---planning cannot be grounded in execution, failures are hard to interpret, and execution traces cannot be reused as learning signals.

When tools become uniformly registered, executable, and traceable, a different regime becomes possible. Agents can plan over realistic action spaces, execute multi-step workflows with reduced stochastic failure, and learn from structured execution traces. Similarly, AI4S models can be evaluated against real scientific tools rather than simplified proxies, enabling more meaningful assessments of practical capability and end-to-end scientific utility.

This framing also clarifies the role of Deploy-Master within broader ecosystem stacks such as Bohrium+SciMaster \cite{zhang2025bohriumscimaster}. In such a stack, Deploy-Master functions as a capability-conversion layer that turns open-source repositories into execution-validated, traceable actions; an orchestration layer (e.g., SciMaster) can then compose these actions into long-horizon workflows executed by a hierarchy of agents, including domain master agents and large numbers of community agents. At scale, repeated executions close a flywheel: traces produced by tool invocation feed back into more reliable deployments (via specification refinement and infrastructure policy updates) and more effective agent behaviors (via better tool selection, avoidance of brittle paths, and targeted remediation for high-value failures). In this sense, deployability is not only a prerequisite for scaling agentic science from isolated demonstrations to systematic experimentation, but also a measurement layer that makes large tool ecosystems learnable and improvable.

\subsection{Limitations and Ongoing Work}
\label{subsec:limitations}

Despite its scale, Deploy-Master does not address all challenges associated with scientific software execution. Several limitations point to directions for ongoing and future work.

\textbf{Hardware heterogeneity.} Many scientific tools depend on specialized accelerators, operating systems, or device drivers that are difficult to abstract within a uniform build pipeline. Supporting such environments requires tighter integration with hardware-aware scheduling, provenance tracking, and execution systems.

\textbf{Distributed and multi-node workflows.} While containerization enables reproducible single-node execution, many scientific applications rely on MPI-based or tightly coupled distributed architectures. Extending deployment support to these settings requires explicit handling of coordination, communication, and resource allocation.

\textbf{Semantic I/O and interoperability.} Minimal executable validation ensures that a tool can run, but does not define standardized input-output contracts or semantic interfaces. Stronger, machine-interpretable I/O specifications are necessary to enable deeper composition, verification, and reasoning across tools.

\textbf{Deeper scenario integration.} Closed-loop systems that connect software execution with laboratory automation, data acquisition, or physical experiments introduce additional constraints around timing, safety, and control that go beyond software deployment alone.

These limitations do not diminish the value of Deploy-Master; they delineate the boundary between making tools runnable and building full-fledged scientific execution ecosystems. Addressing these challenges will require continued work at the intersection of software infrastructure, systems design, and scientific practice.



\bibliography{main}
\bibliographystyle{bst/sn-nature}

\end{document}